\documentclass[epj]{svjour}

\usepackage{graphicx}
\usepackage{ctable}
\usepackage{dcolumn}
\usepackage{bm}
\usepackage{amsmath}
\usepackage{amssymb}
\usepackage{nicefrac}
\usepackage[usenames,dvipsnames,svgnames,table]{xcolor}

\begin{document}

\author{Wolf B. Dapp\inst{1} \and Martin H. M\"user\inst{1,2}
\institute{J\"ulich Supercomputing Centre, Institute for Advanced Simulation, FZ J\"ulich, 52425 J\"ulich, Germany \and
Department of Materials Science and Engineering, Universit\"at des Saarlandes, 66123 Saarbr\"ucken, Germany}
\thanks
{e-mail: \tt{martin.mueser@mx.uni-saarland.de}}
}

\title{
Towards time-dependent, non-equilibrium charge-transfer force fields}
\subtitle{
Contact electrification and history-dependent dissociation limits
}


\abstract 
{
Force fields uniquely assign interatomic forces for a given set of atomic coordinates.
The underlying assumption is that electrons are in their quantum-mechanical ground state or in thermal equilibrium.
However, there is an abundance of cases where this is unjustified
because the system is only locally in equilibrium. 
In particular, the fractional charges of atoms, clusters, or solids
tend to not only depend on atomic positions but also on how the system reached its state.
For example, the charge of an isolated solid -- and thus the forces
between atoms in that solid -- usually depends on
the counterbody with which it has last formed contact. 
Similarly, the charge of an atom, resulting from the dissociation
of a molecule, can differ for different solvents in which 
the dissociation took place.
In this paper we demonstrate that such charge-transfer history effects can 
be accounted for by assigning discrete oxidation states to atoms.
With our method, an atom can donate an integer charge to another, nearby atom to
change its oxidation state as in a redox reaction.
In addition to integer charges, atoms can exchange ``partial charges'' 
which are determined with the split charge equilibration method. 
}

\authorrunning{W. B. Dapp and M. H. M\"user}
\titlerunning{Non-equilibrium charge-transfer force fields}

\maketitle

\section{Introduction}\label{sec:intro}
As an illustration of a system with history-dependent forces consider the following thought experiment:
a neutral gold cluster and a neutral sodium cluster are prepared
separately and placed at a large separation in a UHV chamber.
The two clusters are then moved to close proximity and separated later
in such a way that no atoms have transferred between the clusters.
Since sodium has the smaller work function, the
gold cluster will have picked up electrons from the sodium
cluster after the separation. 
As a result, the two final clusters will carry opposite charge
and thus attract each other through a
long-range Coulomb force, which was not there initially.
This implies that the atomic interactions between -- but also within --
the two clusters differ between the
initial and the final state. This is the case even if we allow for a demon enforcing
the atoms to take identical positions at the beginning and at the end.
Current force fields are intrinsically unable to reflect such
changes in the atomic interactions, because they assign to each 
atomic configuration an unambiguous set of forces. 

Even charge~equilibration (QE) or charge-transfer methods~\cite{rick02,lopes09} 
fail to to describe the electrostatic fields that arise due
to the contact-induced charge transfer.
They determine fractional charges as
single-valued (vector) functions of the instantaneous atomic 
coordinates just like conventional density functional theory provides a unique
electronic charge densities for a given atomic configuration. 
While attempts to incorporate time dependence into density functional 
theory-based approaches to the electronic state are 
thriving~\cite{tddftBook}, 
virtually no attention has been paid to the question how to tackle 
history-dependent fractional charges with classical force fields. 
Being able to do so would not only proof useful to reproduce
non-equilibrium charge transfer processes as they
occur in our thought experiment. It might also become
possible to simulate many other systems and processes 
which are not amenable to simulations based on current force-fields,
or contain too many particles for calculations necessitating
the solution of time-dependent quantum mechanics. 
One such example, discussed in this work, is the dissociation
of an NaCl molecule in solvents, which can result in neutral or
singly-charged dissociation products depending on the
solvent polarity.
Another example, which we discuss in detail in a separate work, 
is the discharge of a small-scale battery~\cite{Dapp12battery}.

The proper description of electron transfer upon contact formation
is a crucial aspect in the all-atom simulation of the discharge of a battery, 
since closing and opening of an electric switch involves the formation
and the breaking of a (mechanical) contact between two solid bodies. 
This can be seen in the following discussion.
Consider a battery which is attached to an electric 
circuit with an Ohmic resistor and an initially open switch.
When the switch is closed, a demon manages to keep all atoms
near their initial positions. 
After some electric current has flown through the external wire,
the switch is opened again.
Although no atom has moved substantially inside the battery, 
the charge distribution, or, more generally speaking,
the electronic state can have changed significantly, because many
electrons, i.e., integer charges, have moved through the wire
and the external Ohmic resistance. 
To summarize, current force-fields or DFT-based methods can predict 
neither the voltage of a Galvanic cell nor its time dependence 
during discharge, because they are intrinsically unable to
describe out-of-equilibrium processes as they occur in our thought 
experiment of the contact formation between two metals, e.g., the closing
of a switch. 

Integer charge transfer between atoms, molecules, clusters, or solids
is not only at the root of contact electrification but also of many 
other processes, in particular of any redox reaction. 
Most redox reactions can be understood in quite simple
terms and experienced chemists can easily make accurate predictions
for the outcome of a redox reaction.
Yet, surprisingly little work has been devoted to
cast this chemical intuition into a quantitative framework that
does not necessitate the solution of the Schr\"odinger or 
related equations. 
However, in principle, it should be possible to design an
approach that does not only reflect chemical intuition
but also describes the energetics of the reactants and the products
of a redox reaction with a precision as high as that of
accurate force fields designed for non-reactive systems.
To date, force fields
require different parameterizations for the products
and the reactants of a redox reaction.
Even force fields designed to describe chemical reactions, such as 
ReaxFF~\cite{Nielson05},
fail to account
for the charge transfer effects discussed in the context of
the contact formation between two clusters and the
discharge of a battery. 
A redox reaction implies a quasi-discontinuous change
of the electronic state~\cite{electronTransferBook}, which cannot be 
captured by any force field
that assumes a continuous evolution of interatomic forces with atomic
coordinates.

In this paper we explore the suggestion that
introducing oxidation states into a charge-transfer approach makes
it possible to capture history dependence in classical force fields 
and to  model (non-equilibrium) redox reactions~\cite{Mueser2012}.
Our intent is to demonstrate a proof of concept rather than to model
a specific system with high accuracy.
In this first attempt, we focus our attention on the thought experiment
discussed at the beginning of the introduction.
In a future publication we will demonstrate that all-atom simulations of batteries
(based on the methodology developed herein) reproduce non-trivial observations,
such as the existence of a voltage plateau during discharge of the battery
for small discharge rates. 

The remainder of this work is organized as follows:
In Section~\ref{sec:method} we summarize the split-charge equilibration
(SQE) model~\cite{NistorEtAl2006}, which we used as a basis for our study.
This includes a discussion of the original method as well as its
extension to \protect{``redoxSQE,''}  in which oxidation states can be
changed. 
In Section~\ref{sec:results} we demonstrate that redoxSQE
indeed captures the time dependence of interactions between molecules
as discussed in our first thought experiment.
Conclusions are drawn in Section~\ref{sec:conclusions}.

\section{Method}\label{sec:method}

\subsection{The split-charge equilibration method}

Charge equilibration methods are based on the minimization of an expression
reflecting the energy as a function of the charge distribution in an atomic 
system~\cite{rick02,lopes09}.
In this work we use the split-charge model~\cite{NistorEtAl2006}:
\begin{equation}
V = \sum_{i} \left( \frac{ \kappa_i } {2} Q_i^2 + \chi_i Q_i \right) +
\sum_{i,j<i} \left( \frac{\kappa_{ij}}{2} q_{ij}^2 + {J_{ij}}{} Q_i Q_j \right).
\label{eq:SQEmodel}
\end{equation}
Here, $Q_i$ is the (possibly fractional) charge of atom $i$. 
Its chemical hardness and electronegativity are denoted by $\kappa_i$ and
$\chi_i$, respectively.
$J_{ij}$ reflects the distance-dependent Coulomb interaction between two atoms,
which we do not screen at short distances, i.e., 
$J_{ij} \equiv 1/r_{ij}$ in this study,
$r_{ij}$ being the distance between two atoms.
The term $q_{ij}$ denotes a partial charge, or ``split charge'', 
donated from atom $j$ to atom $i$. 
It is antisymmetric in the indices, i.e., $q_{ij} = -q_{ji}$.
The net charge of an atom in the formulation proposed in reference~\cite{Mueser2012}
then reads:
\begin{equation}
Q_i = n_i e + \sum_j q_{ij},
\label{eq:charge}
\end{equation}
where $n_i$ is the (formal) oxidation state and $e$ the elementary charge. 
The term $\kappa_{ij}$, which is distance dependent
(and should usually also be environment dependent), is called the 
bond hardness. It is inversely proportional to the
polarizability of the bond connecting two atoms. 
In the condensed phase, the inverse bond hardness associated with
nearest neighbors is proportional to the dielectric 
permittivity~\cite{nistor09prb}.

We refer to the original literature for an in-depth discussion and motivation
of the SQE model~\cite{NistorEtAl2006,nistor09prb,Mueser2012}. 
However, we  summarize some of its properties in this section
and compare it to other charge equilibration approaches.
In contrast to the traditional purely atom-based charge transfer
potentials, i.e., those omitting bond-hardness terms,
the SQE model produces the correct dissociation limit of 
molecules~\cite{NistorEtAl2006,Mueser2012,mathieu07},
the (high-frequency) dielectric permittivity does not automatically
diverge~\cite{nistor09prb},
it shows the correct linear scaling of the polarizability of polymers
with increasing chain length~\cite{warren08,verstraelen09}, as well as
the correct slight reduction of the dipole of alcohols with increasing length of the
fatty tail~\cite{mikulski09}.
Lastly, the chemical hardness of molecules (which one can associate with
the band gap of solids) levels off at a finite value rather than
to become zero for large atom numbers~\cite{Mueser2012}.
In contrast to purely bond-polarizability based models,
i.e., approaches excluding the singly-indexed atomic hardness $\kappa_i$,
SQE shows the correct scaling of oligomers with decreasing chain
length~\cite{warren08}, 
it is not restricted to systems with a dielectric
permittivity just slightly above unity~\cite{Mueser2012,nistor09prb},
the chemical hardness of molecules is positive as it should be, rather than 
negative~\cite{Mueser2012},
and it maintains the concept of atomic hardness, which is well
motivated from
the density functional theory (DFT) formulation of quantum
mechanics~\cite{mortier85jacs,parr83,itskowitz97}.
SQE has also been coupled to bond-order potentials describing 
short-range interactions.
Good values were obtained for dipoles and heat of formation for isolated 
molecules, radial distribution functions of liquids, as well
as reasonable energies of oxygenated diamond surfaces~\cite{knippenberg12}.

The work on charge transfer models by Verstraelen and coworkers 
deserves particular mention.
They found that SQE outperforms previous charge equilibration approaches
in all of their 23 benchmark assessments when applied to a large,
systematically chosen set of organic molecules~\cite{verstraelen09}.
In a later study~\cite{verstraelen12jpcc}, they showed that SQE produces 
good charges and response functions for both isolated silicate 
and related molecules as well as their condensed ceramic phases, by using 
an unambiguous and thus transferable parameterization scheme.
In contrast, atom-based charge transfer approaches suffered from the usual 
superlinear scaling of the polarizability.
In a theoretical work, Verstraelen {\it et al.} 
motivate the
combination of bond hardness and atomic hardness with
DFT-based considerations~\cite{verstraelenMISC}.
In a recent study~\cite{verstraelen12jctc}, they recognized that the original 
SQE model (i.e., the one in which all oxidation states are set to zero) 
fails to predict fractional charges and polarizabilities of zwitterionic 
molecules.
Once charges of (formally charged) fragments in (zwitterionic) molecules
were constrained to be a positive or negative elementary charge plus a 
split charge from the atom to which the fragment was bonded,
fractional charges turned out to compare well to those deduced
from an iterative Hirshfeld partitioning.
Their constraint on 
fragments~\cite{verstraelen12jctc}, called SQE+Q$^0$,
is similar in spirit to redoxSQE,
even if atom-based charge-transfer models cannot be extended to
fragments in a simple, systematic fashion~\cite{Valone11JCTC,Valone11JPCL}. 

\subsection{Bond hardness and bond breaking\label{subsec:kappas}}

In the phenomenological framework of the SQE model, the value of 
the bond-hardness terms classifies the nature of the bond between 
two atoms.
A bond  between two atoms $i$ and $j$ is called metallic
if $\kappa_{ij} = 0$.
When two atoms are not bonded, no split charge can be shared, which 
can be expressed formally with $\kappa_{ij} = \infty$.
If $\kappa_{ij}$ is positive and finite, we classify the bond as 
dielectric.
Of course, conductivity is a collective property rather than a function
of the separation between two atoms, and approaches like ours will
be unable to capture resonance effects, for instance in 
 molecular-scale electronics. 
Yet, we are in a position to set up local rules producing
the correct macroscopic dielectric response functions when properly
parameterized.
We feel that this is progress with regards to routinely-used charge
equilibration approaches (such as the one employed in ReaxFF) which
always yield the response of ideal metals~\cite{nistor09prb,Mueser2012}.

For simplicity, we take the bond hardness as a function solely of the
separation between any two atoms --- or ions, once we have
introduced oxidation states.
This first-order approximation will have to be abandoned for material-specific simulations,
where one might want to design the bond hardness along the lines of
bond-order potentials~\cite{pettifor99}. 
A bond shall be broken, i.e., $\kappa_{ij}$ should diverge, 
if $r_{ij}$ exceeds the threshold value $r_{\rm l}$.
This also implies that atoms $i$ and $j$ no longer share a split charge. 
(In reality, the polarizability between two distant atoms should
never be exactly zero, but decay exponentially as their atomic
orbitals cease to overlap.)
At short separation ($r_{ij} \le r_{\mathrm{s}}$), $\kappa_{ij}$ should move to a value such that
a system composed of dimers at the given separation gives the correct
dielectric response. 

To show the generic features of our model, we use 
a rather basic model, where the bond hardness is either zero (to reflect 
traditional charge equilibration approaches) or a unique function
of distance. 
Making $\kappa_{ij}(r_{ij})$ depend on the oxidation state of the
atoms does not change the property of the model qualitatively.
For a neutral dimer in the redoxSQE description, we set the two
atomic oxidation states to zero, or alternatively,  to $n_{1,2} = \pm 1$
(so that the net charge of the molecule remains zero). 

We adopt the functional form for $\kappa _{ij}$ similar to the one introduced  
by Mathieu~\cite{mathieu07}, who already parameterized bond breaking 
in the split-charge framework.
Specifically, we choose:
\begin{eqnarray}
  \kappa _{ij} = \left\{
        \begin{array}{ll}
                \kappa_{ij}^{\rm (p)}     & r_{ij} \le r_{\mathrm{s}},\\
   \kappa_{ij}^{\rm (p)} +             
\kappa^{(0)}_{ij}
\frac {\left(r_{ij}-r_{\mathrm{s}}\right)^{2}}
   {\left(r_{\mathrm{l}}-r_{ij}\right)^{2}}
                        & r_{\mathrm{s}} < r_{ij} < r_{\mathrm{l}},\\
                \infty  & r_{\mathrm{l}} \le r_{ij}.\\
        \end{array}%
        \right.         
   \label{eq:BondHard}
\end{eqnarray}
Here, $\kappa_{ij}^{(\rm p)}$ is the plateau value, which is zero for metals
and positive for dielectrics, while
$\kappa_{ij}^{(0)}$ is always positive.
We do not claim that the functional form in equation~(\ref{eq:BondHard})
is chemically realistic, 
although Mathieu~\cite{mathieu07} found quite
reasonable trends for fractional charges in the analysis
of the homolysis of a variety of organic molecules.
That work included results for radicals and transition states, 
even though
the calibration was done on equilibrated structures
satisfying the octet rule.

Our diatomic molecule shall be characterized as follows.
Atomic hardness: $\kappa_1 = \kappa_2 = 3.5$ (in arbitrary units,
which in typical cases should correspond roughly to $\rm V/e$), 
and electronegativity: $\chi_{\rm 1,2} = \mp 1.5$, i.e., atom 2 (henceforth
called the anion) is more electronegative than atom 1, which 
one could associate with the cation. 
This choice implies the correct dissociation limit, that is, the energy
required to transfer charge between two distant atoms is positive, 
specifically $\Delta E_{\rm trans} = 0.5$.
Moreover, we choose 
$r_{\rm s} = 0$, $r_{\rm l} = 2$,  $\kappa_{\rm 12}^{\rm (p)} = 0$, and
$\kappa_{\rm 12}^{\rm (0)} = 1/8$. 
The Coulomb interaction reads 
$V_{\rm C} = Q_1 Q_2 / r_{12}$.
The split-charge can be calculated by minimizing equation~(\ref{eq:SQEmodel})
with respect to the split charge $q_{\rm 12}$, i.e.,
\begin{equation}
q_{\rm 12} = \frac{\chi_2+(\kappa_2-J)n_2e-\chi_1-(\kappa_1-J)n_1e}
{\kappa_1 + \kappa_2 + \kappa_{12}-2J}
\label{eq:equilibriumCharge}
\end{equation}
We consider neutral systems, i.e., $n_1 = -n_2$. 
For the discussion of charged systems and molecular hardness, we
refer to reference~\cite{Mueser2012}. 

\begin{figure}[htbp]
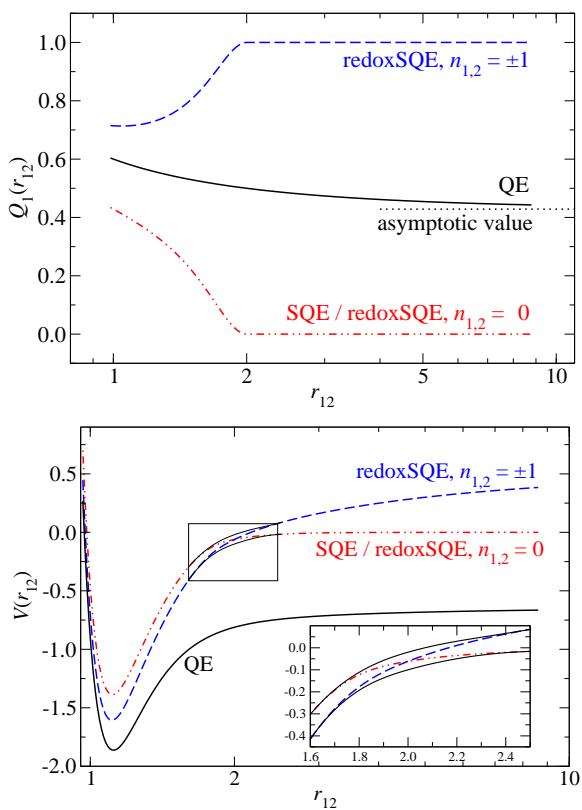

\centering
  \vspace*{5mm} 

 \includegraphics[width=0.85\columnwidth,angle=0]{fig1a.eps}
  \vspace*{2mm}

 \includegraphics[width=0.85\columnwidth,angle=0]{fig1b.eps}
\caption{(Color online) 
In both graphs, two different oxidation state configurations are considered
as well as the conventional charge equilibration approach (QE).
{\bf Top:} 
charge $Q_1$ of the ``cation'' as a function of the 
interatomic distance $r_{12}$ for different computational approaches.
The dotted black line is the charge on the electropositive 
atom predicted by the QE method for an infinite separation.
{\bf Bottom:}
energy $V_{12}$ of the diatomic molecule as a function of the 
interatomic distance.
The short black lines in the bottom graph illustrate
the functional dependence of the energy of the quantum-mechanical
ground state and the first excited state on the interatomic distance.
\label{fig:genericDissoc}
}
\end{figure}

We add a short-range two-body interaction energy
to the split-charge related energies.
For reasons of simplicity we use a standard Lennard-Jones potential:
\begin{equation}
V_{\rm LJ}(r) = 4\epsilon \left\{
\left( \frac{\sigma}{r} \right)^{12}-
\left( \frac{\sigma}{r} \right)^6
\right\},
\end{equation}
where $\epsilon = \sigma = 1$ are chosen. 
Here, a typical value for $\sigma$ would be 2\AA. 
We abstain from adjusting any interaction
parameters for different oxidation states, even if
this level of simplicity is probably not tenable for 
realistic parameterizations. 
The resulting charges on atom 1 and the total energy for different
oxidation states are shown in Figure~\ref{fig:genericDissoc}.

The top panel of Figure~\ref{fig:genericDissoc} illustrates
that conventional QE ($\kappa_{ij} \equiv 0$, choice of $n_1$ irrelevant for neutral
systems)
produces fractional charges on the two atoms even when they are distant. 
At infinite separations, the charge on the electropositive atom
levels off at a finite value $(\chi_2-\chi_1)/(\kappa_1+\kappa_2) \approx 0.43$, 
indicated by the black dotted line.
In redoxSQE however, the two atoms are either neutral, or each one carries an integer charge of 
opposite sign in the dissociation limit,
owing to the diverging bond hardness.

In the bottom part of Figure~\ref{fig:genericDissoc}, we point out
that the two redoxSQE energy curves with different
values for $n_1$ cross. At short separation, the system benefits
if it
 transfers a negative integer charge to the anion, while both
atoms are preferably
neutral in the dissociation limit. 
Qualitatively, the behavior of the two curves is similar to those
obtained from 
full quantum-mechanical potential energies of ground states
and first excited states, except near the transition region.
There, the two curves would not cross 
but the $n_1 = 1$ curve
would smoothly evolve into the $n_1 = 0$ curve and vice versa (as sketched
by the short black lines). 
We note that the crossing of the two levels does not
have to occur at $r_{\rm l}$, but that both existence and location of the
crossing point depend on the choice of parameters. 
Moreover, we note that the asymptotic behavior of the $n_{1,2} = \pm 1$ curve 
for large $r_{12}$ reflects the Coulomb attraction
between two opposite, integer charges as well as the energy $\Delta E_{\rm trans}$
required to ionize atom 1 and donate the electron to atom 2. 

\subsection{Some quantitative analysis for NaCl\label{subsec:NaCl}}

It is beyond the scope of this work to perform a full parameterization
of the method to real materials using quantum-chemical data. However, as demonstration of
the applicability
of the method to a real system, we performed MP2 
calculations of the dissociation of an NaCl dimer in Gaussian~\cite{g03}, 
with 
the basis function set \protect{6-311G+}. These
calculations used the restricted Hartree-Fock method, which enforces
closed electron shells for the dissociation products and causes the 
system upon separation to move onto the ``excited'' energy curve in 
a Landau-Zener picture. 
For NaCl, this implies that $n_{\mathrm{Na},\mathrm{Cl}}=\pm 1$, 
and does not allow for a neutral dissociation, even in vacuum. 
The result is an energy curve very similar to that 
for the charged dissociation of a generic dimer
(dashed blue line in Figure~\ref{fig:genericDissoc}, bottom panel).

\begin{figure}[htbp]
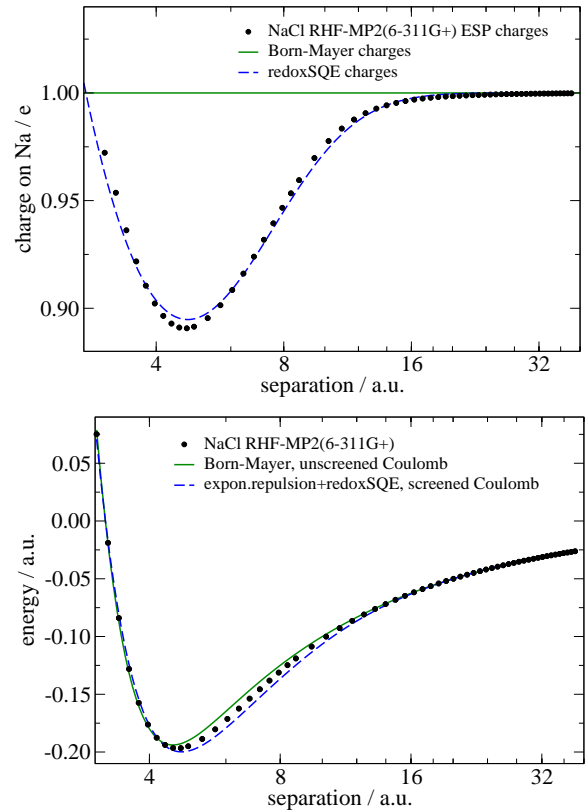

\centering
  \vspace*{5mm} 

 \includegraphics[width=0.85\columnwidth,angle=0]{fig2a.eps}
  \vspace*{2mm}

 \includegraphics[width=0.85\columnwidth,angle=0]{fig2b.eps}
\caption{(Color online) 
\label{fig:MP2data}
{\bf Top:} 
ESP charges on sodium from RHF-MP2(\protect{6-311G+}) calculations with Gaussian 
of the dissociation of NaCl in vacuum, as well the charges
predicted by the redoxSQE model. The Born-Mayer model
uses constant charges $\pm 1$.
The MP2 data also sets the functional form of 
$\kappa_{\mathrm{Na}^{+}\mathrm{Cl}^{-}}$. 
{\bf Bottom:}
Fits to MP2 energy data. 
A simple Born-Mayer potential with an 
unscreened Coulomb interaction and constant charges fits fortuitously well.
The redoxSQE model with a screened Coulomb interaction fits the data 
comparably well. Both models have two free parameters for the
repulsive term. 
In addition the redoxSQE model uses parameters determined from 
MP2 calculations for isolated Na/Na$^{+}$ and Cl/Cl$^{-}$ as well
as the ESP charges of an Na$^+$Cl$^-$ dimer. 
}
\end{figure}

To the MP2 energy curve we fit two models. 
First, a Born-Mayer potential~\cite{BornMayer1931} with a simple 
unscreened (fixed-charge) Coulomb interaction, and two free parameters, 
in the repulsive short-range term. Figure~\ref{fig:MP2data} (bottom panel)
reveals that
this simple, purely two-body potential fits the data surprisingly well. 
This result is fortuitous in parts, because the model uses a
simple unscreened Coulomb interaction.
Allowing for screening did not improve the results.

The second model uses the redoxSQE potential with a screened Coulomb 
interaction for short separations, and self-consistent charges. 
The ionic dissociation necessitates determining hardness and 
electronegativity for the ions 
and not the neutral atoms. 
We obtain the ionization energy of Cl$^{-}$ 
and the electron affinity of Na$^{+}$ from consistent MP2 calculations 
for isolated atom/ions.
Since the electron affinity of the Cl anion is not defined, we have
to work with an effective value. 
We obtain one additional constraint by setting the electronegativity 
difference equal to the screened Coulomb energy close to  where
the electrostatic potential (ESP) 
charges derived from the MP2 calculations cross 
the $Q_{\rm Na}/e = 1$ line. 
These three values are sufficient to gauge the needed single-ion 
parameter combinations
$(\kappa_{{\rm Cl}^{-}} + \kappa_{{\rm Na}^{+}})$ and
$(\chi_{{\rm Cl}^{-}} - \chi_{{\rm Na}^{+}})$.

To reflect screening we replaced the $1/r$ Coulomb potential with
a $\tanh(r/a_{\rm scr})/r$ dependence, which we found to be very similar to
Slater-type orbital screening.
At the present stage of development, we treat $a_{\rm scr}$ as an adjustable
parameter.
We find that, as argued in Section~\ref{subsec:kappas},
a simple 
$\kappa_{\mathrm{Na}^{+}\mathrm{Cl}^{-}} \propto \exp(r/a_{\rm b})$ 
relation is sufficient to reproduce the ESP
charges from the MP2 calculation
at all distances, see top panel of Figure~\ref{fig:MP2data}.

Having determined those parameters, we 
fit the energy curve with two free parameters, namely those in the repulsive 
short-range term, as in the Born-Mayer case. 
The fit of the redoxSQE model to the data is also very good --- the 
error is $\lesssim 4\%$ of the binding energy everywhere.
We note that while Born-Mayer uses only two fit parameters,
we have a total of five adjustable parameters.
However, in our approach, we reproduce the ESP charges, which
are assumed to be constant in Born-Mayer.
Moreover, the resulting values for the fit parameters are physically meaningful.
For example, we obtain a combined ionic hardness of 33~V/$e$,
which is similar in magnitude to the experimental hardness of the sodium cation of
$42$~V/$e$ (the Cl anion contributes very little, because it is chemically soft). 
The bond hardness, evaluated at the NaCl distance in the rocksalt crystal,
is 57~V/$e$.
Given the relation for the high-frequency dielectric permittivity for 
rock salt equation~(27) in reference~\cite{nistor09prb}
this would produce a value of  $\varepsilon_{\rm r} = 2.1$, which
is close to the experimental value of $\varepsilon_{\rm r} = 2.25$.

Note that the charge on the sodium atom resembles the behavior seen in 
Figure~\ref{fig:genericDissoc} (top panel) in that it decreases from very small 
separations (as in the QE case), transitions through a rise at 
intermediate distances, and finally levels off to +1 at large 
separations (the detailed shape of the curve depends on the choice 
of $\kappa_{ij}$, which in the generic case was chosen a divergent
rational function, while $\kappa_{\mathrm{Na}^{+}\mathrm{Cl}^{-}}$ 
increases only exponentially with separation).
A reduction of the Born effective charge with increasing pressure  
is also seen in DFT calculations of solid MgO~\cite{OganovEtAl2003}, from +2 at zero pressure
to $\approx 1.65$ at $\approx 800$~GPa. A similar result can also be derived from 
Raman scattering data on ZnO~\cite{ReparazEtAl2010}, from +1 at zero pressure
to $\approx 0.96$ at $\approx 5.5$~GPa. Such reductions can only be 
achieved with redoxSQE and the inclusion of oxidation states, 
as conventional SQE (including QE) automatically predicts a 
monotonic \textit{increase} in charge upon compression.
This is because the increasing Coulomb interaction upon compression
always increases the split charge when the reference state consists
of neutral atoms. 

In future work, we will carry out a more detailed study on the 
transferability of the redoxSQE parameters. 

\subsection{Assigning oxidation states}

Before addressing how to design dynamics for the transition between
the $n_{1,2} = \pm 1$ and the $n_{1,2} = 0$ surface,
we wish to comment on the asymptotic behavior of interatomic forces
near $r_{\rm l}$.
Since $q$ minimizes the potential energy, the interatomic
force reads, in the generic two-body problem considered here:
\begin{equation}
{\bf F}({r_{12}}) = 
{\bf F}_{\rm LJ}({r_{12}}) +
{\bf F}_{\rm C}({r_{12}}) -
q_{12}^2 \nabla \kappa_{12}(r_{12}).
\end{equation}
In addition to the Lennard-Jones force, 
${\bf F}_{\rm LJ}$, and Coulomb force between (point-) charges,
${\bf F}_{\rm C}$, an additional term arises as a result of the
distance dependence in the bond hardness.
We disregard possible
environment-dependent corrections to atomic hardness and electronegativity. 
If we now assume that $\kappa_{ij}$ diverges with a power law
$1/{(r_{\rm l}- r_{12})^{\alpha_{\rm l}}}$, the absolute value of the  
additional force term scales as 
$q_{12}^2/(r_{\rm l}-r_{12})^{\alpha_{\rm l}+1}$.
The split charges themselves scale 
as $q_{ij} \propto 1/\kappa_{ij}$ when $r_{ij}$ approaches $r_{\rm l}$.
Thus, the extra contribution to the force originating from the bond-hardness terms
becomes
\begin{equation}
- q_{12}^2 \nabla \kappa_{12}(r_{12}) \propto  
(r_{\rm l}-r_{12})^{\alpha_{\rm l}-1}.
\end{equation}
Choosing $0 < \alpha_{\rm l} < 1$ would lead to an integrable 
singularity in the force, so that the bond could be broken with
finite energy. 
However, a numerical implementation has the high risk of leading
to unstable simulations. 
For $\alpha_{\rm l} = 1$, the force is discontinuous at $r_{\rm l}$,
which implies that  the usually used 
symplectic (Verlet) integration algorithms would no longer
be strictly energy conserving (this would be roughly equivalent to cutting off potentials at finite
distances without smoothing the potential). 
Our choice of $\alpha_{\rm l} = 2$ leads to a discontinuity only in the 
first derivative
of the force. This means that the second derivative is divergent at isolated
points so that symplectic integrators might still show very small drifts,
which should nevertheless be unnoticeable in practice. 

The behavior for the discontinuity in the force-distance curve at the 
short-range cutoff can be discussed in similar terms as that at large 
distances. 
Now, $q$ does not necessarily tend to zero and must therefore
be assumed to be finite at $r = r_{\rm s}$. 
Thus, assuming that the ``excess'' hardness,
i.e., $\kappa_{12}-\kappa_{12}^{(0)}$  increases
with $(r-r_{\rm s})^{\alpha_{\rm s}}\Theta(r-r_{\rm s})$,
one obtains a discontinuous force
at $r_{\rm s}$
for $\alpha_{\rm s} = 1$ and a discontinuity in the first
derivative of the force for $\alpha_{\rm s} = 2$. Here, $\Theta$ is 
the Heaviside step function.

Simulating the dissociation of molecules, clusters, or solids
necessitates rules for the change of the oxidation state of 
individual atoms -- or at least of fragments.
The feature of the redoxSQE model that we explore in this work
is that it allows for integer charge transfer (ICT) besides the exchange
of partial charges across bonds between two atoms.
This section explains our implementation of such ICTs. 
First, however, we motivate the rules by performing an asymptotic analysis
of a charge transfer process, which is similar in spirit to the
thought experiments discussed in the introduction.  

\begin{figure}
 \centering
 \includegraphics[width=0.95\columnwidth,angle=0]{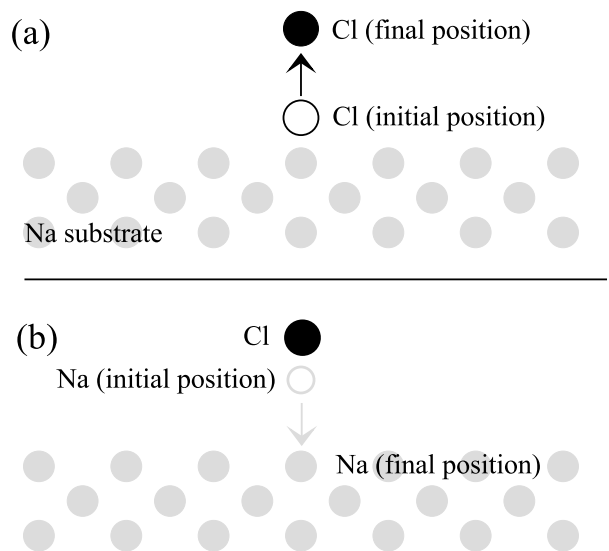}
\caption{
Schematic illustration of the history dependence of atomic charges.
The final nuclear configurations in (a) and (b) are identical.
However, in (a), the chlorine atom has initially resided on the surface
of the metal cluster so that its final charge will be a negative
elementary charge. In (b), an NaCl molecule is dissociated first leading
to two neutral atoms. When the Na is introduced into the bulk substrate,
the Cl is too distant to receive an electron from the bulk metal 
on relevant time scales. 
The different outcome despite identical final positions 
arises because the electron affinity of chlorine
exceeds the work function of metallic sodium but is smaller than the
ionization energy of atomic sodium. 
\label{fig:schematic}
}
\end{figure}

Consider a chlorine atom close to a surface of metallic sodium in UHV,
as illustrated in Figure~\ref{fig:schematic}(a).
Initially the Cl atom is close to the Na metal and is moved adiabatically
to a distant position.
Straight from the onset, 
the sodium cluster will have donated an electron to chlorine.
At no point in time will the chlorine be able to
reduce its energy by donating its extra electron to the sodium,
because the electron affinity of chlorine exceeds the work function
of solid sodium.
Thus, the final charge of the Cl atom will be -1 and that of
the Na cluster +1. 

In contrast, if we reach the final configuration through the process shown in
Figure~\ref{fig:schematic}(b), the chlorine will remain neutral 
on experimentally relevant time scales. The reason is that an (adiabatic)
dissociation of the NaCl dimer leads to two neutral atoms, because
the ionization energy of Na exceeds the electron affinity of Cl.
Thus, the final states for the two cases differ. We stress that we did not assume any 
``extreme'' conditions, such as high shear rates or pressures,
as they typically occur at microscopic contact points 
between two rubbing solids.

Determining the precise rates of electron transfer reactions 
is a mature field of research~\cite{electronTransferBook}.
The transfer of electrons between atoms can occur with non-negligible
probabilities on time scales shorter than a molecular dynamics time step
when two atoms are not too distant from one another.
Once they have moved away from one another, the rates
decrease exponentially with separation so that they
eventually become negligible even over extended periods of time. 
In many cases, such as the desorption/dissociation of a Cl atom from sodium 
discussed in this section,
the outcome 
does not
depend sensitively on the time when this transfer has occurred.
In our current example, the precise electron tunneling rate would 
only matter if the molecular dissociation in case (b) occurred at
an ``intermediate'' distance from the surface. 
In all other cases, it would suffice to have an algorithm 
producing approximate average rates for the electron transfer process.

To mimic electron tunneling dynamics we proceed as follows:
at each time step, we select all ``relevant'' dielectric bonds.
As mentioned before, a bond is called dielectric if two atoms are
sufficiently close together to share a split charge, but not close
enough to have a vanishing bond hardness.
Which pairs of atoms are classified as sharing a relevant bond
depends on the problem of interest.
In the current example, it would only be the bond between
the chlorine atom and one sodium atom if their distance is small ---
or the bonds to several atoms at the bulk sodium surface.
In the contact formation between two solids discussed in the
introduction, relevant bonds would be the ones between pairs 
of atoms from different metal clusters. 
Non-relevant dielectric bonds are those where electron hopping can be 
neglected, for example, between two atoms belonging to the same dielectric
material, which we assume to be non-conducting, or two atoms far apart
(with a separation $>r_{\rm l}$).
In general, the number of relevant bonds should be 
$<gNZ$, where $N$ is the number of atoms, $Z$ the average coordination
number, and $c^2<g<c$.
Here, $c$ denotes the fraction of atoms capable of redox reactions. 
One would expect $g \propto c^2$ if the redox-active atoms are perfectly
dissolved in a redox-inert solvent, 
while $g \lesssim c$ if they do not dissolve at all. 

Once a bond has been identified as relevant, we examine certain 
additional conditions to determine whether to attempt an ICT. 
Some conditions that could 
stifle an ICT are physical constraints. An example is 
the ability of a donor to give an electron (e.g., H$^{+}$ 
cannot donate an electron), or the possibility for an acceptor 
to receive it (e.g., it is exceedingly unlikely that Cl$^{-}$ 
accepts an electron). Second, in order to reflect electron 
transfer probabilities, we require that the product of time step 
size and estimated transfer rate is greater than a random 
number between zero and one. (In chemically realistic 
simulations, the transfer rate must be determined from quantum-
mechanical calculations~\cite{Tully1990}.) 
In addition, in order to alleviate a 
bias introduced by the order in which we query bonds, we draw 
an additional random number and only proceed with the trial ICT 
if it exceeds some threshold 
(for instance $0.9$, to attempt an ICT only every tenth MD time step). 
An easy way to avoid bias altogether is to pick the trial-ICT 
bonds at random from the pool of relevant bonds. 
For the examples studied herein, this turned out unnecessary, 
which we ascertain from the observation that the symmetry of the
problem was maintained.

If all conditions are met, we make a trial ICT.
This involves changing the oxidation state of the two bonded 
partners, incrementing one, and decrementing the other by 1, with the
direction of transfer set by the sign of the current split charge between 
the two partners. We then re-equilibrate all split charges within some 
cutoff radius $R^{\mathrm{opt}}$. This approach implicitly assumes that the electronic 
response to a discontinuous change in the electric field, as it is brought 
about by the ICT, is faster than the inertial response of the atoms, which 
is traced by an MD time step. 

In our current, developmental implementation $R^{\mathrm{opt}} = \infty$, which 
means that all split charges are updated in each trial move.
This makes each individual trial ICT an expensive ${\cal O}(N^2)$ operation
and thus, a full sweep through the system is ${\cal O}(gZN^3)$. 
Assuming that \textit{all} split charges react 
(instantaneously) to a changed electric field, is similar 
to saying that {\it all molecular  orbitals of the system will be
modified during an (electron transfer) reaction}~\cite{electronTransferBook}.
However, it should often be possible to turn the trial ICTs into 
local operations with bounded errors~\cite{Mueser2012}, by choosing a finite
cutoff radius. The reason is 
that the change in the split charges induced by the redox move 
decreases very quickly with increasing
distance to the location of the electron transfer.

After the equilibration we calculate the total potential energy of the 
system, except for the charge-independent short-range terms. Generally, 
atoms that changed their oxidation states would have differing short-range 
interactions with their neighbors as well. For instance, the chlorine 
anion is significantly larger than a neutral chlorine atom. For now, we 
neglect this change. If the energy after the change of oxidation state and 
subsequent split-charge equilibration is equal to or below what it was 
before, the ICT is accepted, otherwise it is rejected. In other words: for 
each atomic configuration, knowledge of the set of oxidation states allows 
one to deduce on which (Landau-Zener-type) energy surface the system is 
currently evolving. An ICT allows to hop from one surface to another, 
lower one. 
A natural alternative to this minimal energy procedure, which keeps the 
electrons close to their ground states, would be to apply the Metropolis 
algorithm. In case of a rejection, we undo the changes in 
oxidation state and restore all split charges to their values before the trial ICT.
If a move is accepted, the next relevant bond is examined, on the basis of the 
modified charge distribution in the system. When this process is exhausted, the
new charge distribution is used for the force calculation for the next MD step.

Note that the described procedure is not energy conserving. We implicitly assume 
that the excess energy liberated by the electron transfer escapes as radiation. 
It is not difficult to design a modified ICT that is radiation-free, for 
instance by adding the released energy to the kinetic energy of the two 
participating atoms (and modify the acceptance criterion of a trial ICT
accordingly), for instance, as in reference~\cite{Tully1990}. 
We leave this extension for future work.

ICTs across metallic bonds ($\kappa_{ij} \equiv 0$) are unnecessary, 
because backflow of the split charge $q_{ij}$ exactly compensates
the integer charge transfer. 
Atomic charges and total energies are unaffected in this case, so that every trial
move would be accepted, but no interatomic force would change.
Rather than assigning oxidation states to individual atoms sharing a metallic
bond,
it would be more meaningful to assign them to connected metal clusters.
However, in many cases, such as the ones discussed in this section,
the overhead of constructing clusters can be avoided.

\section{Case studies}\label{sec:results}

\subsection{Contact formation between metals}
The mechanism of contact electrification between two metals is well 
established~\cite{Lowell80}.
Electrons transfer from the metal having the smaller work function to 
that with the larger one. 
The resulting charged solids then have a reduced difference of
their work functions and the charge transfer eventually comes to a stop. 
Of course, the amount of transferred charge is affected by electrostatics,
i.e., by the total capacitance of the two metals and 
by the rate at which the two metals are pulled apart. 
Thus, while the {\it amount} of transferred charge may be difficult to predict 
theoretically, the {\it direction} of charge flow between initially 
neutral {\it metals}
is entirely determined by their work functions.
Therefore, metals can be arranged in 
a well-defined fashion into the triboelectric
series~\cite{Lowell80,Diaz04}, which lists materials closer to the top (bottom) 
the more it has the propensity to have attracted negative (positive) charge 
after having touched or rubbed against a counter body.

\begin{figure}
 \centering
 \includegraphics[width=0.95\columnwidth,angle=0]{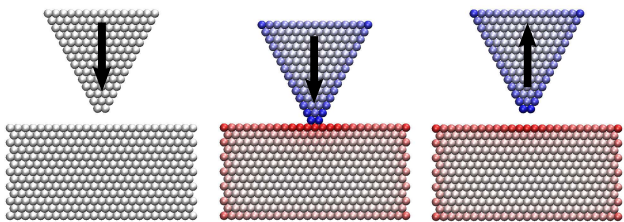}
\caption{Visualization of the contact electrification of a metal tip and a metal
substrate. 
Red and blue indicate, respectively, the amount of negative and positive charge per atom.
Initially neither solid is charged. When they are brought to close proximity,
charge can transfer between them such that negative charge flows from the metal
with the small work function to that with the large work function.
After the solids have been separated, no more charge can flow. 
\label{fig:Metals}
}
\end{figure}

Figure~\ref{fig:Metals} demonstrates that our model reproduces the qualitative
features of charge transfer between metals. 
It shows a tip which is brought into close contact with a substrate and then
separated again.
The clusters are initially neutral; however, during contact, charge can 
transfer, and remain (to some extent) on the clusters after separation.
The charge distribution mainly lives 
on the surface of the clusters, 
as expected for metals, because this minimizes the
(repulsive) electrostatic energy.

In the simulations for Figure~\ref{fig:Metals}, 
the center of mass of the tip was moved by a small 
increment at each time step, and all atoms within the tip
were kept fixed relative to its center-of-mass.
Within each solid the bond hardness was set to zero.
Bonds between atoms belonging to different solids turned dielectric 
--- and could thus share a split charge --- when
their separation was below the threshold $r_{\rm l}$.
For the calculations shown here, a trial ITC was made 
in each dielectric bond with a trial rate
inversely proportional to the bond hardness.
The electron tunneling rate therefore also has this proportionality, 
i.e., it decreases quadratically with increasing separation between 
the clusters, and vanishes at $r_{ij} = r_{\rm l}$.
One can certainly envision other rules for the attempt frequency, 
for instance an exponential decrease with distance, which 
corresponds more closely to an exponentially decreasing
electron tunneling rate.
However, the details of such rules do not affect the results in a qualitative
fashion, which is why we abstained from ``fine tuning'' the parameters.

\begin{figure}
\centering
\includegraphics[width=0.85\columnwidth,angle=0]{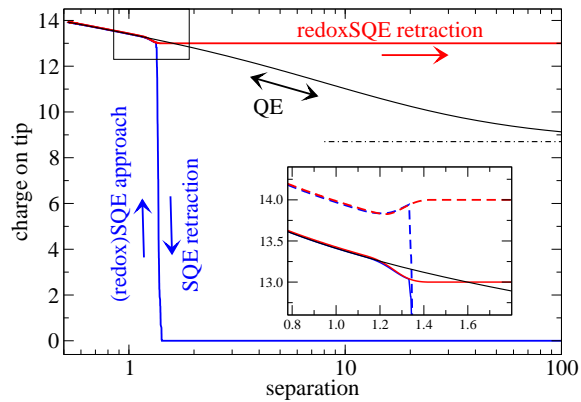}
\caption{\label{fig:qTransferMetal}
Charge of the electropositive metal cluster from Figure~\ref{fig:Metals} 
as a function of separation. The blue line represents the charge
curve using the original SQE method during both approach 
\textit{and} retraction. Using the redoxSQE formalism, the 
charge follows the blue curve during approach, but the red one during 
retraction. The black line reflects the results from a traditional 
charge equilibration method.
The dot-dashed black line corresponds to the charge transfer in the
QE model when tip and substrate are infinitely far apart. 
{\bf Inset:} blow-up of the redoxSQE charge-distance curve.
The dashed lines represent a simulation in which the electronegativity
difference between tip and substrate atoms was increased by 
$\approx 10\%$ with respect to the original data. 
}
\end{figure}

We repeated the same calculation with a pure QE model,
in which (split) charge can flow freely between all atoms.
As a consequence, 
the charge distribution is a unique function of the 
distance between the two parts, 
in a similar fashion as in the formation and breaking of a diatomic
molecule in Figure~\ref{fig:genericDissoc}.
The results for the metallic tip-substrate system are
shown in Figure~\ref{fig:qTransferMetal}.
The charge transfer is driven by
the electronegativity difference and opposed by the generation of
an electrostatic field around each metal cluster, i.e., by the
(inverse) capacitance  of the two metal pieces, which is predominantly 
a function of the shape of tip and substrate, and, for small clusters,
also a function of the atomic hardness. 

The inset of Figure~\ref{fig:qTransferMetal} reveals that the charge on the
electropositive tip does not necessarily decrease 
monotonically with the separation between tip and substrate.
The final charge tends to be the closest integer value
to the charge at a separation $r \lesssim r_{\rm l}$,
which can result in a rounding-up of the charge as in the
$n_1 = 1$ case in the top panels of Figure~\ref{fig:genericDissoc} and \ref{fig:MP2data}.
Conversely, if the next integer charge is more easily reached by
rounding down, the charge behaves more like in the
$n_{1,2} =0$ case of Figure~\ref{fig:genericDissoc}. 

\subsection{Contact formation between dielectrics}

The charge transfer between dielectrics is much less well
understood than that between metals.
Competing views are that rubbing or other contact-dynamics induced
charging occurs {\it predominantly} through electron 
transfer (potentially via gap surface states)~\cite{mate08book}, 
proton transfer~\cite{Diaz04}, or the exchange of  
hydroxide or other ions~\cite{McCarthy08}.
Unlike metals, dielectrics cannot be arranged in a unambiguous fashion
into a triboelectric series.
In some cases, it is even possible to arrange materials
in a cyclic triboelectric series. For example, in the list
{\{}glass, zinc, silk, filter paper, cotton, glass\}
each material attracts negative charge during rubbing from that standing
to its immediate left~\cite{harper98}.

The existence of cyclic triboelectric series implies that tribocharging 
between dielectrics cannot generally be rationalized in terms of 
unique or effective electron affinities and ionization energies of the
involved (free-standing) dielectrics. 
However, it may be premature to conclude that the existence
of cyclic triboelectric series can only be explained by
ion transport as argued previously~\cite{McCarthy08}.
Local pressures at points of intimate mechanical contact can be very 
high for brief moments, i.e., shorter than the time needed to nucleate 
a dislocation.
Consequently, 
local stresses can reach the theoretical yield stress of the underlying 
materials for times sufficiently long for
electron transfer reactions to take place.
During the time interval when pressures are extreme,
hybridization changes and/or the generation of 
charged molecules can occur in a tribological 
contact~\cite{mosey05sci,mosey05prb}, and
dramatically 
altered charge transfer characteristics may result.

The model used in this work is too crude to reflect hybridization effects,
surface gap states or other effects whose descriptions would 
necessitate sophisticated parameterization schemes.
However, this deficiency is not intrinsic to the idea of assigning 
oxidation states to atoms, but rather a consequence of restricting
the parameterization to two-body terms.  
In our approach, positive charge tends to flow from the
electronegative to the electropositive material, and, as far as mono-atomic
solids are concerned, the model would result in a well-defined 
triboelectric series.

\begin{figure}
\centering
 \includegraphics[width=0.95\columnwidth,angle=0]{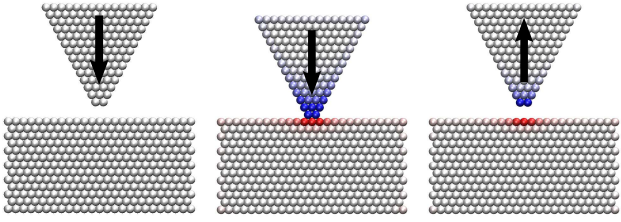}
\caption{\label{fig:Diel}
Similar to Figure~\ref{fig:Metals}.
However, this time, the bonds within each solid are modeled
as dielectric.
This prevents the charge from delocalizing significantly. 
\label{fig:DielCharge}
}
\end{figure}

Apart from this limitation, the \protect{redoxSQE} approach introduced in this
work can reproduce qualitative
features of  the charge transfer between two dielectrics, as shown 
in Figure~\ref{fig:Diel}.
Charge can be exchanged between two contacting solids or clusters.
Unlike those placed onto a metal, charges being donated to a dielectric 
do not spread out over the surface but remain pinned near the point of contact.
This behavior would also be found in real systems, except that electrons would 
be allowed to hop on long time scales.
This, in principle, could have been 
incorporated in the simulations as well.

\begin{figure}
\centering
\includegraphics[width=0.85\columnwidth,angle=0]{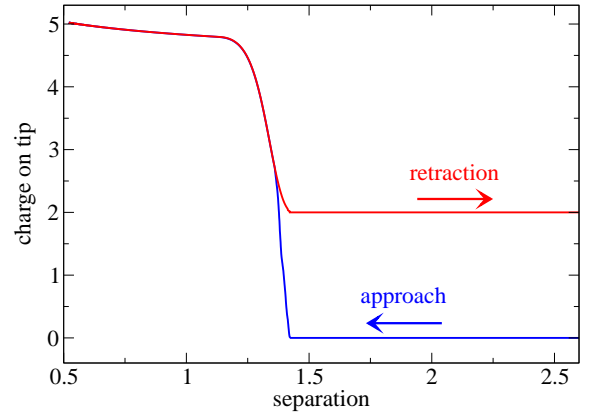}
\caption{\label{fig:qTransferDiel}
Charge of the electropositive dielectric tip from Figure~\ref{fig:Diel} 
as a function of separation during approach and retraction in the redoxSQE model.
}
\end{figure}

Although the atomic hardnesses for the contact formation between two dielectrics
chosen the same as those used in the simulation of metal clusters, 
the total
amount of transferred charge is much less, which can be 
seen when comparing Figure~\ref{fig:qTransferDiel} to Figure~\ref{fig:qTransferMetal}.
Since ICTs are only attempted between the two front atoms of the tip with
the closest substrate atoms, and not passed on further within the cluster,
the charge that is ultimately transferred is restricted
to two elementary charges for the given set up.
In comparison, the number of transferred elementary charges is 13 for the 
two studied metal clusters.
Another difference is that polarization charge in the dielectric is located near
the atoms with non-zero oxidation number, while it is spread much more evenly
across the metal surface. 

\subsection{Semi-quantitative considerations}

The two examples of contact electrification 
discussed above do not pertain to any particular material.
However, they reflect reality in the sense that collectivity 
induces the possibility to transfer integer charges between (SQE) materials,
even if the atoms constituting the electropositive material have 
an ionization energy $I$
exceeding the electron affinity $A$ of the atoms composing the
counter body. 
In this section, we expand the analysis of how collectivity
affects the values for $I$ and $A$ within the redoxSQE model 
--- and thus redox characteristics ---
in two specific systems, namely adsorbed atoms on metals and
water molecules near an NaCl dimer.

In a pioneering work almost 80 years ago, Gurney~\cite{gurney35} discussed 
why the work function of tungsten can be be reduced through a few
alkaline earth atoms.
He argued that this observation cannot be explained in a classical
picture since ionizing, say, a single Ca atom costs 
$I_{\rm Ca} \approx 6.1$~eV,
which exceeds the work function of tungsten
$W_{\rm W} \approx 4.5$~eV. 
Yet, redoxSQE reproduces the observed experimental trend. 
Given that all bonds are metallic (the argument remains unchanged if the
redoxSQE bond between Ca and W has a large but finite polarizability), 
charge can be shared among different
(isolated) Ca atoms on the surface. This diminishes the effect of the
atomic hardness, and electronegativity becomes the relevant 
parameter, i.e., $\chi_{\rm Ca} = 1$~V. 
Since values for $\chi$ differ only slightly between alkaline and alkaline
earth metals, in contrast to their ionization energies, redoxSQE finds
a similar reduction in $W_{\rm W}$ for alkaline and alkaline earth
adsorbates, in agreement with the observations reported in 
reference~\cite{gurney35}. 

For the adsorbed metals, ionization potentials were reduced because
new dielectric or metallic bonds allowed to spread the charge.
External potentials can also achieve such a reduction, as we now discuss.
Without hydration, a NaCl dimer dissociates in its ground state
into two neutral atoms because $I_{\rm Na} \approx 5$~eV is greater
than the electron affinity of chlorine, $A_{\rm Cl} \approx 3.6$~eV. 
In order to estimate the effect of hydration for the dissociated dimer,
let us assume that electrostatic interactions are dominant at a separation
of $r = 2.5$~{\AA} between a water molecule and an ion of integer charge.
The electrostatic potential of a dipole is 
$ \Phi = {\bf d}\cdot{\bf r}/(4\pi\epsilon_0 r^3)$. 
Thus the energy gain for a perfectly oriented dimer ${\bf d}$ is
$\Delta E = d r/(8\pi\epsilon_0 r^2)$.
Inserting $d = 1.85$~debye (the value for an isolated H$_2$O molecule) yields
$\Delta E \approx 0.4$~eV for singly charged ions. 
Thus, approximately two water molecules per ion are sufficient
to topple the energy balance toward a charged dissociation limit. 
We conclude that redoxSQE, once well parameterized, would find 
--- and remember! --- that rock salt molecules
decompose into neutral atoms in an inert-gas atmosphere
but as ions in a sufficiently polar solvent, although the
latter screens the direct Coulomb interaction between the ions. 

\section{Discussion and Conclusions}\label{sec:conclusions}

In this paper, we showed that introducing formal oxidation states
to the split charge model allows one to reproduce various generic
charge transfer features that occur during the contact formation and breaking
of two solid clusters. 
The main idea behind the approach is that oxidation states of atoms
can only change via a redox reaction involving 
two or more nearby atoms. 
This way partial charges are neither automatically constant nor a well-defined
function of the instantaneous configuration but may be history dependent.

An important many-body effect implicitly contained in the model is that 
integer charge transfer can occur between two dielectrics of different
electron affinity. 
This happens even though breaking (adiabatically) any molecule consisting
of two (stable) atoms in an inert environment automatically results in 
two neutral atoms, at least for realistic values of their
atomic hardnesses and electronegativities.
The reason for this behavior lies in a reduced ionization energy $I$
for collective systems in comparison to isolated atoms.
In a future publication, we will demonstrate that the methodology explored
in this work moreover reproduces a variety of generic features that occur
during the discharge and the aging of batteries~\cite{Dapp12battery}.
No currently popular force field can achieve this, because they all assume 
a one-to-one correspondence between atomic coordinates and interatomic 
forces. 
Time-dependent density functional theory approaches are likewise
likely to fail to produce such features due to the serious constraints
on particle numbers and simulation times. 

Our case study reproducing the generic features of contact-induced charging 
between two dielectrics assumes implicitly that electric current is effected by the
passage of integer charges, i.e., electrons, from one solid to the other. 
We nevertheless do not intend to take sides in the discussion of what charge types
are mainly responsible for charge transfer~\cite{Diaz04,McCarthy08}.
Protons, hydroxides, or other ions might be equally or even more
important than electrons.
Our purpose is to provide a formalism with which this issue can be addressed
in terms of many-atom, force-field based simulations.
This includes the ability to ascertain whether an atom leaves a solid or a cluster
as a neutral atom or as an ion. 
Moreover, the redoxSQE method allows for the (local) electron or ion
exchange between 
stoichiometrically similar materials, as long as local properties 
materials are distinct, e.g., because of chemical heterogeneity
or different local radii of curvature. 
Thus, there is a chance to identify reasons for the frequently observed
charge transfer between chemically identical materials. 
Of course, recognizing the main mechanism of contact electrification 
for a specific
system requires a realistic and chemistry-specific parameterization.

It is encouraging that the original SQE formalism  reproduces fractional charges 
and polarizabilities quite accurately for systems in which atoms have a 
well-defined oxidation 
state~\cite{NistorEtAl2006,mathieu07,mikulski09,verstraelen09,verstraelen12jpcc,verstraelen12jctc,knippenberg12}.
Moreover, it has 
been shown that the reactant and the product of a redox reaction,
namely terminally blocked and zwitterionic penta-alanine,
can be described quite accurately if the functional groups are subjected to 
constraints~\cite{verstraelen12jctc}.
The effects of the constraints are very similar to those induced by assigning
discrete oxidation states to atoms. 
We are thus confident that it is possible to describe products and reactants
of redox reactions reasonably well within one unified framework, 
although it might be necessary to define (effective) 
chemical hardness values for all relevant oxidation states.

What still has to be achieved, however, is a way to assign reasonable rates 
for the redox reactions, in particular near the transition state.
In this context it needs to be mentioned that our classical description  of
an isolated diatomic molecule produces a zero gap between ground state
and first excited state at the transition state.
This violates the non-crossing rule of quantum mechanics~\cite{Neumann29}.

As for any coarse-graining effort, 
the important question is whether the information lost 
is relevant to the problem of interest. 
To answer this question, we first note that coarse-graining --- as we do it here 
implicitly from a full quantum-mechanical description of the electronic structure 
to partial charges and oxidation states --- must entail loss of information,
unless the full description is not optimal. 
As a benefit, larger systems and longer time scales become accessible. 
Second, most experiments studying contact charging between 
dielectrics~\cite{Diaz04,McCarthy08}, in particular
those where electron hopping in the dielectrics are negligible, find
that the rate of withdrawal of the surfaces barely affects the final charge.
This observation can be rationalized quite easily by noticing that the tunneling 
rate of electrons or the transfer rates of ions decrease roughly exponentially with the
distance between two solids.
Last but not least, in the all-atom battery simulations that will be presented in a future publication,
we find that both cutoff $r_{\rm l}$ and the rate of trial integer-charge transfer
moves do not change results within the statistical error bars. 
We therefore conclude that it should indeed be possible to parameterize \protect{redoxSQE} 
for many
applications in such a way that the outcome of the simulations produce the correct
ensemble average, e.g., accurate distributions of charge transfer between
colliding clusters or rubbing solids.

\bigskip
\bigskip
\footnotesize
We thank the J\"ulich Supercomputing Centre for computing time
and Razvan Nistor and Ling Ti Kong for useful discussions. Also, we thank Michael 
Springborg and Jorge Vargas for assistance with the MP2 calculations.


\bibliographystyle{unsrt}
\bibliography{./bibliography,./newMMbiblio}
\end{document}